\documentclass[aps,prl,showpacs,preprintnumbers,amsmath,amssymb,twocolumn]{revtex4}
\usepackage{setspace}
\usepackage{graphicx}
\begin{document}

\title {An Exact Expression for a Flat Connection on the Complement of a Torus
Knot}

\author{V. V. Sreedhar\footnote{\sl sreedhar@cmi.ac.in}}  

\affiliation{Chennai Mathematical Institute, Plot H1, SIPCOT IT 
Park, Siruseri, Kelambakkam Post, Chennai 603103, India}

\begin{abstract}
Simple physics ideas are used to derive an exact expression for a 
flat connection on the complement of a torus knot. The result is of
some importance in the context of constructing representations of the 
knot group -- a topological invariant of the knot. It is also a step 
forward in the direction of obtaining a generalisation of the 
Aharonov-Bohm effect in which charged particles moving through 
force-free regions are scattered by impenetrable, knotted solenoids. 

\end{abstract}

\pacs{03.65.Ta, 02.10.Kn, 03.50De}

\maketitle

This paper deals with the problem of deriving a flat connection on 
the complement of a torus knot. The motivation for undertaking this 
exercise is two-fold. First, any exact expression is interesting in 
its own right. Besides, in the present case, the result is useful in 
constructing holonomies and hence the representations of the knot 
group -- a well-known topological invariant of the knot \cite{ma}. 
Second, the result has a direct physical application. It is 
a first step towards generalising the seminal work of Aharonov and 
Bohm, on the quantum mechanical scattering of charged particles moving 
through force-free fields \cite{ab}, to a situation in which the 
impenetrable solenoid is knotted \cite{abknot}. The key idea that is 
used in this paper to obtain the result relies on modelling a knot 
with wires and solenoids carrying steady currents. The fields associated 
with these objects can then be calculated in a straightforward manner by 
standard methods of classical electrodynamics.  

As a prelude to the calculation, it is useful to briefly recapitulate 
some relevant, but well-known, mathematical facts about knots 
\cite{knotes}. A knot $K$ is a closed, oriented, loop of string in 
${\bf R}^3$. It is defined by the map $K: S^{1} \rightarrow {\bf R}^3$. 
Two knots $K_1, K_2 \in {\bf R}^3$ are equivalent if there exists an 
orientation-preserving homeomorphism $h:{\bf R}^3 \rightarrow {\bf R}^3$ 
such that $h(K_1) = K_2$.  

Let $X$ be the complement, ${\bf R}^3 - K$, of the knot $K$. This is a 
path-connected (non-compact) 3-manifold. The knot determines the complement. 
Clearly, equivalent knots have homeomorphic complements. 

The knot group is, by definition, the fundamental group $\pi_1(X)$.
Since complements of equivalent knots are homeomorphic, their fundamental 
groups are isomorphic.

A converse theorem \cite{gl} states that knots are determined by their 
complements. In other words, two knots having homeomorphic complements 
are equivalent. A second converse theorem \cite{whitten} states that 
if two prime knots have isomorphic groups, their complements are 
homeomorphic which, by virtue of the first theorem, implies that the 
two knots are equivalent. Hence, the knot group determines the knot.
It is a topological invariant associated with the knot. In general, knot 
groups are nonabelian. The knot group of a trefoil knot, for example, is 
the braid group on three strings, ${\bf B}_3$. 

To put the above statements into perspective, let us consider a two-dimensional
analogue. The space relevant to the standard Aharonov-Bohm effect is a plane 
with a hole. The fundamental group of this space is the (abelian) additive 
group of integers, ${\bf Z}$.  As is well-known, an exact expression, in 
Cartesian coordinates, for a flat connection on this space is given by the 
formula $\vec A (x,y,z) = {\Phi\over 2\pi (x^2 + y^2)} (-y, x, 0)$, where 
$\Phi$ is the flux through the hole. What is the corresponding expression 
for a flat connection on the complement of a knot? The present exercise 
answers this question.    

Consider a small tubular neighbourhood of the knot with cross-sectional 
radius $\epsilon$. Imagine a densely packed winding around it, with a wire 
carrying a constant current $i$ per winding. Let $n$ be the number of windings 
per unit length, and $dl^{'}$ an infinitesimal line element at $\vec r^{'}$
along the tube. This knotted solenoid is a simple, albeit nontrivial, 
generalisation of the more familiar current distributions used in the study of 
the Aharonov-Bohm effect {\it viz.} the solenoidal and toroidal distributions 
in which the winding is done around a cylinder and a torus respectively. From 
basic magnetostatics, the winding produces a magnetic field which has support 
only inside the knotted tube. In the complement of the knot, the vector 
potential (connection) is non-zero but the magnetic field vanishes imposing 
the flatness condition. 

The brute-force method to calculate the vector potential at any point in 
the complement of the knotted solenoid follows the standard technique of 
adding contributions from the individual loops. This leads to an integral 
(over the length of the knot) of elliptic integrals (contribution from the 
individual windings). This answer is not very illuminating: A simplification 
by way of a reduction in the number of integrals is desirable.  

Strictly speaking, the winding produces a component of the current along
the knot which produces a non-zero magnetic field outside the knotted tube. 
The resulting field can be cancelled by passing an appropriate current 
through the axis of the knotted tube, in the opposite direction to the 
winding. This in turn produces an additional contribution to the vector 
potential which should be accounted for.  

That is not all: The above current distribution also produces a field due 
to contact terms which we call the `knot moments' -- analogous to the toroidal 
moment, also known as the anapole \cite{zeld} in the case of a toroidal 
winding. Such fields typically have their support only in the source region;
nevertheless, they produce a vector potential outside the sources. This 
produces a further contribution to the result.

All the three caveats mentioned above can be circumvented by the following
simple expedient: We let the radius $\epsilon$ of the knot tube to be small.
Each winding can then be approximated by a magnetic dipole of 
strength $\mu$, at the position $\vec r^{'}$, pointing in the direction
of the tangent to the knot at $\vec r^{'}$. In this limit, the knotted 
solenoid reduces to a collection of magnetic moments (magnets) which, 
for a trefoil knot, are aligned as shown in the figure.

\begin{figure} [<h>]
\includegraphics[scale=0.3]{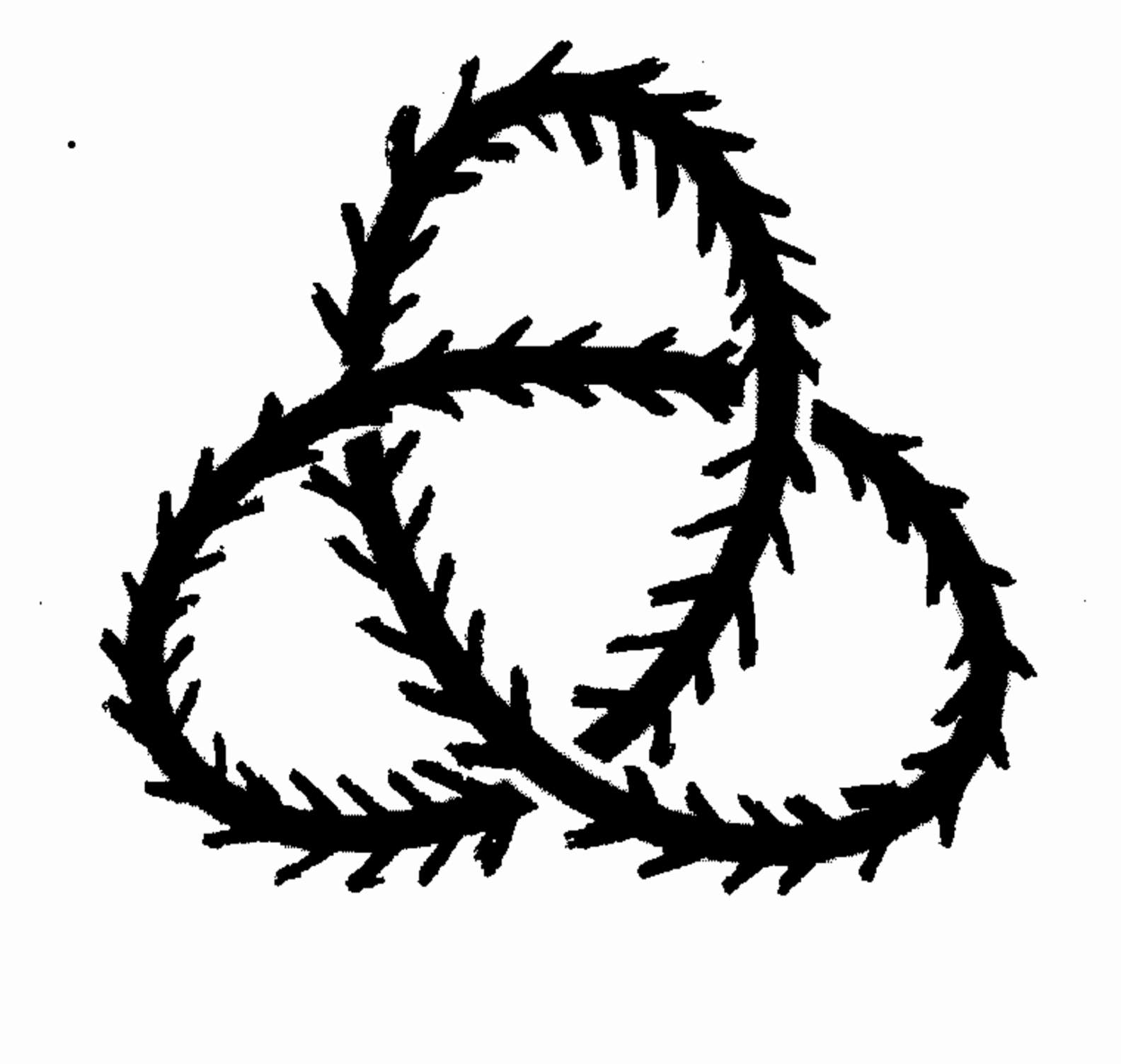}
\caption{Alignment of Magnetic Moments}
\label{Moments}
\end{figure}

Each moment contributes a vector potential equal to ${\vec \mu\times \vec 
R\over R^3}$ at the point $\vec r$, where $\vec R = \vec r - \vec r^{'}$ 
and $R =  \mid \vec R \mid $. The total vector potential is then obtained by 
integrating over $\vec r^{'}$. Thus 
\begin {equation}
\vec A (\vec r) = \int_K dl^{'} \vec m (\vec r^{'})\times {\vec R\over
R^3}
\end{equation} 
where $\vec m$ is the magnetic dipole moment density {\it i.e.} magnetic
dipole moment per unit length. This expression for the vector potential
is identical to the expression for the magnetic field produced by a knotted 
wire carrying steady current, given by the Biot-Savart law, when $\vec m$ 
is replaced by current. The vector potential is given by an expression 
usually reserved for the magnetic field because the former (in the Coulomb
gauge) satisfies the same equations as the latter in the magnetostatics 
limit in the region of interest {\it viz.} the source-free complement of 
the knot. Assuming that the knot tube is of uniform cross-sectional area, and 
defining the Hertz potential $\vec H$ by 
\begin {equation}
\vec H (\vec r) = \mid\vec m\mid \int_K {\vec dl^{'} \over R}
\end{equation} 
the expression for the flat connection (1) can be obtained by 
taking the curl of $\vec H$, $\vec A = \vec\nabla\times \vec H$.  
Equation (2) is deceptively simple since, although the integration is 
over a one-dimensional (filamentary) current, the non-trivial embedding
of the knot makes the description manifestly three-dimensional. The 
key to evaluating this integral is to reduce the description to a 
one-dimensional integral. This simplification is easily effected for a 
class of knots called torus knots by using toroidal coordinates.   
 
A $(p,q)$ torus knot can be obtained by considering a closed path that 
loops around one of the cycles of a putative torus $p$ times, while 
looping around the other cycle $q$ times, $p,q$ being relatively 
prime integers. The toroidal coordinates are denoted by $ 0\leq\eta<\infty,
 ~~ -\pi < \theta \leq \pi , ~~ 0 \leq \phi <2 \pi $. Given a toroidal
surface of major radius $R$ and minor radius $d$, we introduce a 
dimensional parameter $a$, defined by $a^2 = R^2 - d^2$, and a dimensionless
parameter $\eta_0$, defined by $\eta_0 = {\hbox{cosh}}^{-1}(R/d)$.  
The equation $\eta^{'}$ = constant, say $\eta_0$, defines a toroidal 
surface. The combination $R/d$ is called the aspect ratio. 
Clearly, larger $\eta_0$ corresponds to smaller thickness of the putative 
torus. Further, since we are interested in torus knots, we impose the 
constraint: $p\theta^{'} + q\phi^{'} = 0$, $p$ and $q$ being relatively 
prime integers. It follows that $\theta^{'} \rightarrow \theta^{'} + 2\pi 
q \Rightarrow \phi^{'} \rightarrow \phi^{'} - 2\pi p$ {\it i.e.} as we 
complete $q$ cycles in the $\theta$ direction, we are forced to complete 
$p$ cycles in the $\phi$ direction -- as required. These constraints on the 
source coordinates effectively reduce the calculation to a one-dimensional 
problem. 

It should be mentioned that in a different context, namely in the study 
of helical windings on a tokomak, this problem has been studied in great 
detail \cite{bhadra}. For our purposes the results of \cite{mirin} are 
more suitable, and we use them in what follows. 

The toroidal coordinates are related to the usual Cartesian coordinates 
by the equations $x = {a{\hbox{sinh}}\eta{\hbox{cos}}\phi\over 
({\hbox{cosh}}\eta - {\hbox{cos}} \theta )},~ y = {a{\hbox{sinh}}\eta{\hbox
{sin}}\phi\over ({\hbox{cosh}}\eta - {\hbox{cos}} \theta )},~ 
z = {a{\hbox{sin}}\theta\over ({\hbox{cosh}}\eta - {\hbox{cos}}
\theta )}$. The metric coefficients are $h_1 = h_2 = {a\over 
({\hbox{cosh}}\eta - {\hbox{cos}}\theta)}, ~ h_3 = h_1{\hbox{sinh}}\eta $
and the volume element is $dV = {a^3{\hbox{sinh}}\eta \over ({\hbox{cosh}}
\eta - {\hbox{cos}}\theta)^3}$. These results are useful in expressing
the Cartesian components of the Hertz potential in terms of the toroidal 
coordinates. 

Likewise, the Green's functions, for $\eta^{'} > \eta$ and $\eta^{'} < \eta$ 
respectively, are readily expanded in toroidal harmonics as follows:
\begin{widetext}
\begin{equation}
{1\over R} = {1\over a\pi}\sqrt{ ({\hbox{cosh}}\eta^{'} - {\hbox{cos}}
\theta^{'})({\hbox{cosh}}\eta - {\hbox{cos}}\theta )} 
\sum_{m,n=0}^\infty \epsilon_m\epsilon_n (-1)^m 
{\hbox{cos}}m(\phi -\phi^{'})  
{\hbox{cos}}n(\theta -\theta^{'}) 
\{ {P^{-m}_{n-1/2} ({\hbox{cosh}}\eta) Q^m_{n-1/2} ({\hbox{cosh}}
\eta^{'}) 
\atop
 P^{-m}_{n-1/2} ({\hbox{cosh}}\eta^{'}) Q^m_{n-1/2} ({\hbox{cosh}}\eta) 
} 
\end{equation}
\end{widetext}
where the Neumann factor $\epsilon_n$ is equal to $1$ for $n=0$ and 
$2$ for $n\neq 0$, and $P^{-m}_{n-1/2}$ and  $Q^m_{n-1/2}$ are generalised 
associated Legendre functions of the first and second kind with half-integral 
degree. Note that since $\eta^{'} = \eta_0$ defines the putative torus on 
which the knot (of moments) winds, both $\eta > \eta^{'}$ and $\eta < \eta^{'}$
are coordinates in the complement of the knot. Hence both solutions are
of interest. Substituting the above results in equation (2), 
the Cartesian components of the Hertz potential can be calculated by first 
expanding them in toroidal coordinates as follows: 
$H^i (\eta, \theta, \phi) = \sum_{n=0}^\infty \sum_{m=0}^\infty H^i_{nm}$ 
where
\begin{widetext}
\begin{equation}
H^i_{nm} = \sqrt{{\hbox {cosh}}\eta - {\hbox{cos}}\theta} D_{nm} Q^m_{n-1/2}
({\hbox{cosh}}\eta) 
\lbrack 
\alpha^i_{nm}~ {\hbox{cos}}m\phi {\hbox{cos}}n\theta 
+  \beta^i_{nm}~ {\hbox{cos}}m\phi {\hbox{sin}}n\theta 
+  \gamma^i_{nm}~ {\hbox{sin}}m\phi {\hbox{cos}}n\theta 
+  \delta^i_{nm}~ {\hbox{sin}}m\phi {\hbox{sin}}n\theta 
\rbrack
\end{equation}
\end{widetext}
for $\eta > \eta^{'}$ and $D_{nm} = \epsilon_n\epsilon_m (-1)^m/a\pi$. 
The coefficients $\alpha, \beta,
\gamma, $ and $\delta$ are obtained by integrating over the source currents 
and hence contain the information about the knot. The expression for 
$\alpha^i_{nm}$, for example, is given by 
\begin{widetext}
\begin{equation}
\alpha^i_{nm} = \int {a^3 J_i\over ({\hbox{cosh}}\eta^{'} - {\hbox{cos}}
\theta^{'})^{5/2}} P^{-m}_{n-1/2} ({\hbox{cosh}}\eta^{'}) {\hbox{sinh}}\eta^{'} 
{\hbox{cos}}m\phi^{'} {\hbox{cos}}n\theta^{'} d\eta^{'} d\theta^{'} d\phi^{'}
\end{equation}
\end{widetext}
The $\beta^i_{nm} $ and $\gamma^i_{nm}$ are obtained from $\alpha^i_{nm}$ 
by the changes ${\hbox{cos}}n\theta^{'} \rightarrow {\hbox{sin}}n\theta^{'}$
and ${\hbox{cos}}m\phi^{'} \rightarrow {\hbox{sin}}m\phi^{'}$ respectively. 
The $\delta^i_{nm}$ is obtained from $\alpha^i_{nm}$ by making both the 
changes mentioned above. The results for $\eta < \eta^{'}$, are simply obtained
by exchanging the roles of $P^{-m}_{n-1/2}$ and $Q^m_{n-1/2}$.  
The $J_i$ stand for the Cartesian components of the current density and 
can be obtained from the corresponding toroidal components by a change of 
coordinates {\it viz.} $J_i = \sum_\alpha \gamma_{\alpha i}J_\alpha$. 
Here $i = x,y,z; ~\alpha = \eta, \theta, \phi$ and $\gamma_{\alpha i} = 
{1\over h_\alpha} {\partial\xi_i\over \partial\xi_\alpha}$. The non-vanishing 
toroidal components of the current are given by
\begin {equation}
J_\phi \propto {\hbox{cos}}\sigma \delta (\eta^{'} - \eta_0) 
\delta (p\theta^{'} + q\phi^{'}) (h_1 h_2)^{-1}
\end{equation}
and 
\begin {equation}
J_\theta \propto {\hbox{sin}}\sigma \delta (\eta^{'} - \eta_0) 
\delta (p\theta^{'} + q\phi^{'}) (h_1 h_3)^{-1}
\end{equation}
In the above, $\sigma$ is the pitch angle of the knot which winds around the
torus, with respect to the azimuthal direction. The delta function in 
$\eta^{'}$
specifies the putative torus around which the knot (of moments)
winds. The angular delta function enforces the knot constraint 
$ p\theta^{'} + q\phi^{'} = 0$. 
The irksome denominator in (5) can be 
tamed by using the identity \cite{hobson}
\begin{equation}
[{\hbox{cosh}}\eta - {\hbox {cos}}\theta]^{-{1\over 2}} = 
{\sqrt 2\over \pi} \sum_{n=0}^\infty \epsilon_n Q_{n-1/2}({\hbox{cosh}}\eta )
{\hbox{cos}}n\theta
\end{equation} 
Substituting the above results in (5) and performing the integrals gives, 
for the $x$-component of $\alpha$, 
\begin{widetext}
\begin{equation}
\alpha^x_{nm} = \sqrt {2\over 1 + \Lambda_0^2}({a\over\pi}){\hbox{sinh}}\eta_0 
P^{-m}_{n-1/2} ({\hbox{cosh}}\eta_0) \sum_{r=0}^\infty \epsilon_r\bigl[
(-2\Lambda_0)Q^{'}_{r-1/2} ({\hbox{cosh}}\eta_0){\cal I}^{(\alpha)x}_{rmn} 
(p,q) + 
Q_{r-1/2}({\hbox{cosh}}\eta_0) {\cal J}^{(\alpha)x}_{rmn} (p,q) \bigr] 
\end{equation}
\end{widetext}
where $\Lambda_0 = {\hbox{tan}}\sigma = -{(q/p)\over{\hbox{sinh}}\eta_0}$,
and the prime on $Q$ denotes a derivative of $Q$ with respect to $\eta_0$.

The ${\cal I}^{(\alpha)x}$ and ${\cal J}^{(\alpha)x}$ are given by simple 
integrals over elementary trigonometric functions, and are easily evaluated.
Equation (9) holds also for $\alpha_{nm}^i$ with corresponding integrals 
${\cal I}^{(\alpha)i}$ and ${\cal J}^{(\alpha)i}$. For the $z$-component, 
$\alpha_{nm}^z$, however, the ${\cal I}$ and ${\cal J}$ integrals pick up 
the following extra multiplicative factors respectively.
\begin{equation}
{\cal I} \rightarrow {\hbox{sinh}}\eta_0 {\cal I}~~~{\hbox{and}}~~~
{\cal J} \rightarrow \Lambda_0 {\hbox{coth}}\eta_0 {\cal J}
\end{equation}   
The expressions for the ${\cal I}$ and ${\cal J}$ integrals for the other 
cases {\it viz.} $\beta, \gamma, \delta$ can be worked out similarly. 

In evaluating the above integrals, it is useful to let $\lambda = -q/p$, 
and define 
\begin{equation*}
a_1 = \lambda (1-n+r), ~a_2 = \lambda (1+n-r),\\
\end{equation*}
\begin{equation}
a_3 = \lambda (1-n-r),~a_4 = \lambda (1+n+r)
\end{equation}
and
\begin{equation}
b_1 = \lambda (n-r), ~b_2 = \lambda (n+r)
\end{equation}
With these definitions, the values of the integrals for the various cases 
are given below.
\begin{equation}
{\cal I}^x_\alpha =  {1\over 16}\sum_{i=1}^4 f(a_i)({\hbox{cos}}2\pi a_i - 1)
\end{equation}
\begin{equation}
{\cal I}^y_\alpha =  {1\over 16}\sum_{i=1}^4 g(a_i){\hbox{sin}}2\pi a_i 
\end{equation}
\begin{equation}
{\cal I}^z_\alpha = -{1\over 4}\sum_{i=1}^2 h(b_i){\hbox{sin}}2\pi b_i 
\end{equation}
\begin{equation}
{\cal J}^x_\alpha = {1\over 8}\sum_{i=1}^2 g(b_i)({\hbox{cos}}2\pi b_i -1) 
\end{equation}
\begin{equation}
{\cal J}^y_\alpha = {1\over 8}\sum_{i=1}^2 f(b_i){\hbox{sin}}2\pi b_i 
\end{equation}
\begin{equation}
{\cal J}^z_\alpha = {1\over 4}\sum_{i=1}^2 h(b_i){\hbox{sin}}2\pi b_i 
\end{equation}
\begin{equation}
{\cal I}^x_\beta = {1\over 16}\sum_{i=1}^4 f(a_i){\hbox{sin}}2\pi a_i 
\end{equation}
\begin{equation}
{\cal I}^y_\beta = -{1\over 16}\sum_{i=1}^4 g(a_i)({\hbox{cos}}2\pi a_i -1) 
\end{equation}
\begin{equation}
{\cal I}^z_\beta = {1\over 4}\sum_{i=1}^2 h(b_i)({\hbox{cos}}2\pi b_i -1) 
\end{equation}
\begin{equation}
{\cal J}^x_\beta = {1\over 8}\sum_{i=1}^2 g(b_i){\hbox{sin}}2\pi b_i 
\end{equation}
\begin{equation}
{\cal J}^y_\beta = -{1\over 8}\sum_{i=1}^2 f(b_i)({\hbox{cos}}2\pi b_i -1) 
\end{equation}
\begin{equation}
{\cal J}^z_\beta = -{1\over 4}\sum_{i=1}^2 h(b_i)({\hbox{cos}}2\pi b_i -1) 
\end{equation}
\begin{equation}
{\cal I}^x_\gamma = {1\over 16}\sum_{i=1}^4 k(a_i){\hbox{sin}}2\pi a_i 
\end{equation}
\begin{equation}
{\cal I}^y_\gamma = {1\over 16}\sum_{i=1}^4 l(a_i)({\hbox{cos}}2\pi a_i -1) 
\end{equation}
\begin{equation}
{\cal I}^z_\gamma = {1\over 4}\sum_{i=1}^2 \tilde{h}(b_i)({\hbox{cos}}2\pi 
b_i -1) 
\end{equation}
\begin{equation}
{\cal J}^x_\gamma = -{1\over 8}\sum_{i=1}^2 l(b_i){\hbox{sin}}2\pi b_i 
\end{equation}
\begin{equation}
{\cal J}^y_\gamma = -{1\over 8}\sum_{i=1}^2 k(b_i)({\hbox{cos}}2\pi b_i - 1) 
\end{equation}
\begin{equation}
{\cal J}^z_\gamma = -{1\over 4}\sum_{i=1}^2 \tilde{h}(b_i)({\hbox{cos}}2\pi 
b_i -1) 
\end{equation}
\begin{equation}
{\cal I}^x_\delta = -{1\over 16}\sum_{i=1}^4 k(a_i)({\hbox{cos}}2\pi a_i -1) 
\end{equation}
\begin{equation}
{\cal I}^y_\delta = {1\over 16}\sum_{i=1}^4 l(a_i){\hbox{sin}}2\pi a_i  
\end{equation}
\begin{equation}
{\cal I}^z_\delta = {1\over 4}\sum_{i=1}^2 {\tilde h}(b_i){\hbox{sin}}2\pi b_i 
\end{equation}
\begin{equation}
{\cal J}^x_\delta = {1\over 8}\sum_{i=1}^4 l(a_i)({\hbox{cos}}2\pi b_i -1) 
\end{equation}
\begin{equation}
{\cal J}^y_\delta = -{1\over 8}\sum_{i=1}^2 k(b_i){\hbox{sin}}2\pi b_i  
\end{equation}
\begin{equation}
{\cal J}^z_\delta = -{1\over 4}\sum_{i=1}^2{\tilde h}(b_i){\hbox{sin}}2\pi b_i  
\end{equation}
In the above equations $f(x),~h(x),~l(x)$ are odd functions and $g(x),
~\tilde{h}(x)$ and $k(x)$ are even functions defined by the following 
combinations 
\begin{widetext}
\begin{equation}
f(x) = {1\over x+m+1} + {1\over x+m-1} + {1\over x-m+1} +
{1\over x-m-1} 
\end{equation}
\begin{equation}
g(x) = {1\over 1-x-m} + {1\over 1+x+m} + {1\over 1+x-m} +
{1\over 1-x+m} 
\end{equation}
\begin{equation}
h(x) = {1\over x+m} + {1\over x-m},\quad \tilde{h}(x) = {1\over x+m} - 
{1\over x-m} 
\end{equation}
\begin{equation}
k(x) = {1\over m+x+1} + {1\over m-x+1} + {1\over m+x-1} +
{1\over m-x-1} 
\end{equation}
\begin{equation}
l(x) = {1\over m+1-x} + {1\over m-1+x} + {1\over -m+x+1} +
{1\over -m-x-1} 
\end{equation}
\end{widetext}
This completes the derivation of the Hertz potential (4) for an 
arbitrary torus knot. 

To develop some insight into the result, we now specialise to the case
of a trefoil knot. As mentioned earlier, a trefoil is a (2,3) torus knot. 
Substituting $\lambda = -p/q = -3/2$ in the expressions (11) - (36),
the coefficients $\alpha,\beta,\gamma,\delta$ can be evaluated 
explicitly and we find that only two of them are non-zero for each
component. Plugging the results into equation (4), the Hertz potential
is found to have the following form:
\begin{widetext}
\begin{equation}
H^x = \sum_{m,n =0}^\infty H^x_{nm} = \sqrt{{\hbox {cosh}}\eta - {\hbox{cos}}
\theta} \sum_{m,n =0}^\infty D_{nm} Q^m_{n-1/2} ({\hbox{cosh}}\eta) 
\lbrack 
X_{nm}(\eta_0)~ {\hbox{cos}}m\phi {\hbox{cos}}n\theta 
+ \tilde {X}_{nm}(\eta_0 )~ {\hbox{sin}}m\phi {\hbox{sin}}n\theta 
\rbrack
\end{equation}
\begin{equation}
H^y = \sum_{m,n =0}^\infty H^y_{nm} = \sqrt{{\hbox {cosh}}\eta - {\hbox{cos}}
\theta}\sum_{m,n=0}^\infty D_{nm} Q^m_{n-1/2} ({\hbox{cosh}}\eta) 
\lbrack Y_{nm}(\eta_0)~ {\hbox{cos}}m\phi {\hbox{sin}}n\theta 
+ \tilde {Y}_{nm}(\eta_0 )~ {\hbox{sin}}m\phi {\hbox{cos}}n\theta 
\rbrack
\end{equation}
\begin{equation}
H^z = \sum_{m,n=0}^\infty H^z_{nm} = \sqrt{{\hbox {cosh}}\eta - {\hbox{cos}}
\theta} \sum_{m,n =0}^\infty D_{nm} Q^m_{n-1/2} ({\hbox{cosh}}\eta) \lbrack 
Z_{nm}(\eta_0)~ {\hbox{cos}}m\phi {\hbox{sin}}n \theta + \tilde {Z}_{nm}
(\eta_0 )~ {\hbox{sin}}m\phi {\hbox{cos}}n\theta \rbrack
\end{equation}
\end{widetext}
for $\eta > \eta^{'}$ and $D_{nm} = \epsilon_n\epsilon_m (-1)^m/a\pi$. 
The coveted expression for the flat connection can then be obtained by 
taking the curl of the Hertz potential given by the following generic
expressions for the curl of a vector field in toroidal coordinates. 
\begin{widetext}
\begin{equation}
A_\eta  =  {({\hbox{cosh}}\eta - {\hbox{cos}}\theta)\over a}
\bigl[ -{\hbox{sin}}\phi {\partial H_x\over\partial\theta} +
{\hbox{cos}}\phi {\partial H_y\over\partial\theta}\bigr]
+ { ({\hbox{sin}}\theta)\over a} \lbrack {\hbox{cos}}\phi {\partial H_x\over
\partial\phi} + {\hbox{sin}}\phi {\partial H_y\over\partial\phi} \rbrack 
+ {(1 - {\hbox{cosh}}\eta{\hbox{cos}}\theta )\over a{\hbox{sinh}}\eta}
{\partial H_z\over\partial \phi}
\end{equation}
\begin{equation}
A_\theta  =  {(1 - {\hbox{cosh}}\eta{\hbox{cos}}\theta )
\over a{\hbox{sinh}}\eta}
\bigl[ {\hbox{cos}}\phi {\partial H_x\over\partial\phi} +
{\hbox{sin}}\phi {\partial H_y\over\partial\phi}\bigr]
- { ({\hbox{sin}}\theta)\over a}{\partial H_z\over\partial \phi}
-  {({\hbox{cosh}}\eta - {\hbox{cos}}\theta)\over a}
\bigl[ -{\hbox{sin}}\phi {\partial H_x\over\partial\eta} +
{\hbox{cos}}\phi {\partial H_y\over\partial\eta}\bigr]
\end{equation}
\begin{equation}
A_\phi =  
- {(1 - {\hbox{cosh}}\eta{\hbox{cos}}\theta )
\over a}
\bigl[ {\hbox{cos}}\phi {\partial H_x\over\partial\theta} +
{\hbox{sin}}\phi {\partial H_y\over\partial\theta} + 
{\partial H_z\over\partial\eta}\bigr]
- { ({\hbox{sinh}}\eta{\hbox{sin}}\theta)\over a}
\bigl[ {\hbox{cos}}\phi {\partial H_x\over\partial\eta} +
{\hbox{sin}}\phi {\partial H_y\over\partial\eta}  -
{\partial H_z\over\partial\theta}\bigr]
\end{equation}
\end{widetext}

To summarise, we have derived an exact expression for a flat connection 
on the complement of a torus knot. The derivation relies on successfully 
mapping the mathematical problem into a simple physics problem in 
magnetostatics. We conclude by noting that the results for the Hertz 
potential and the flat connection can be taken over to represent the vector 
potential and the magnetic field respectively, produced by a knotted wire
of the same size and shape carrying steady current. 

A few other problems -- some readily doable, and some harder -- naturally 
come to mind. First, it may be of interest in engineering, for some special 
purposes, to design knotted antennae \cite{werner}. This would require going 
beyond the magnetostatic limit discussed in this paper to time-dependent 
situations. Second, it would be interesting to study multipole expansions of 
knot currents in general and, in particular, construct the generalisation 
of the toroidal moment (anapole) for knotted solenoids. Third, it would be of 
considerable mathematical interest to work out analogous results on the 
complement of a figure-eight knot (which is not a torus knot). This is an 
example of a three-dimensional hyperbolic space and plays an important role 
in Thurston's geometrisation programme \cite{thurston}. Next, some effort 
needs to be devoted towards generalising the ideas to the nonabelian case. 
Finally, it would be of considerable interest to study the diffraction and
scattering effects of knotted solenoids on electrons, both theoretically 
and experimentally; thus generalising the work initiated by Ehrenberg, Siday, 
Aharonov and Bohm. We hope to return to these issues elsewhere.

\begin{acknowledgements}

I thank G. Krishnaswami, R. Loganayagam, M. K. Vemuri, R. Nityananda,
and T. R. Ramadas for discussions, and R. Vijaya for making some literature 
available. I am also grateful to an anonymous referee for some useful 
suggestions. 

\end{acknowledgements}

\end{document}